\documentclass[aps,prl,showpacs,twocolumn,amsmath,amssymb,superscriptaddress]{revtex4}
\usepackage{graphicx}
\usepackage{bm}
\usepackage{amssymb}
\usepackage{amsmath}
\usepackage{latexsym}
\usepackage{color}
\newcommand{\ind}[1]{_{#1}}    
\newcommand{\indrm}[1]{_{\mathrm {#1}}}    

\newcommand{\dai}[1]{\Delta \theta_{\ind{#1}}^{({\mathrm s})}}   
\newcommand{\cddw}{CD\filter DW}   
\newcommand{\mono}{\Sigma}   
\newcommand{\filter}{F} 
\newcommand{\cw}{CFW}   
\newcommand{\pathlength}{{L}}   
\newcommand{\CT}{Czerny-Turner}   
\newcommand{\dirate}{{\mathcal D}}   
\newcommand{\diratecum}{{\mathcal D}_{\indrm{\mono }}}   
\newcommand{\mmixed}{$(+,-,-,+)$}

\begin{document}  
\title{Enhanced X-Ray Angular Dispersion and X-ray Spectrographs with Resolving Power Beyond 10$^{8}$}


\author{Yuri Shvyd'ko}\email{shvydko@aps.anl.gov}  \affiliation{Advanced Photon Source,   Argonne National Laboratory, Argonne, Illinois 60439, USA} 


\begin{abstract} 
  Spectrograph is an optical device that is used to disperse photons
  of different energies $E$ into distinct directions and space
  locations, and to take a snapshot of the whole spectrum of photon
  energies with a spatially sensitive photon detector.  Substantial
  advantage of a spectrograph over an ordinary spectral analyzer, is
  its ability to deal with many photon energies simultaneously, thus
  reducing exposure time per spectrum considerably. To realize a
  spectrograph, dispersing elements with large angular dispersion rate
  are required.  Here we show, on the example of CDW x-ray optics
  \cite{Shvydko-SB,SLK06,ShSS11}, that multi-crystal arrangements may
  feature cumulative angular dispersion rates more than an order of
  magnitude larger than those attainable in single Bragg reflections.
  This makes, first, hard x-ray spectrographs feasible, and, secondly,
  a resolving power beyond $E/\Delta E \gtrsim 10^{8}$ achievable.

\end{abstract}


\maketitle


Spectrograph is capable of measuring simultaneously a spectrum of
photon energies $E$, and therefore is an optical device more desirable
than an ordinary spectral analyzer, that measures energy of a single
photon at a time. Spectrographs are especially wanted when long
exposures (data collecting time) are required. This is always the
case, if a very high resolving power is involved and the light sources
are not especially bright. High-resolution spectroscopies in hard
x-ray regime, which are used to study atomic dynamics and electronic
excitations, is the realm where spectrographs would be extremely
favorable due to mentioned constrains. 
Here we show that angular dispersion of hard x-rays can be enhanced by
more than an order of magnitude in multi-crystal arrangements, and
this effect can be used to realize spectrographs in hard x-ray
regime. Notably, the spectrographs become feasible with a resolution
power of $E/\Delta E \gtrsim 10^{8}$, what may
advance significantly research using high-resolution x-ray spectroscopies, in particular inelastic x-ray scattering.

\begin{figure}
\setlength{\unitlength}{\textwidth}
\begin{picture}(1,0.30)(0,0)
\put(0.0,0.00){\includegraphics[width=0.5\textwidth]{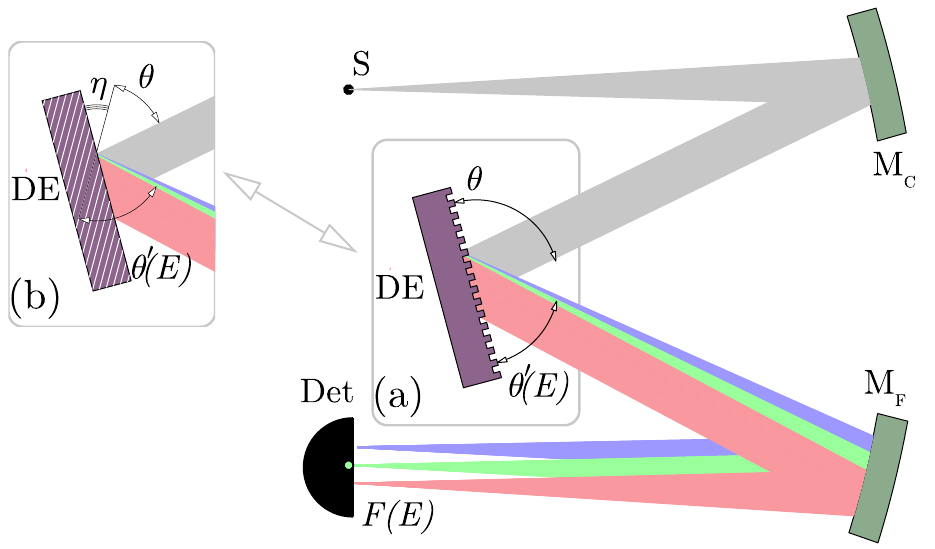}}
\end{picture}
\caption{Scheme of the \CT\ type \cite{CT30} spectrograph with
  diffraction grating (a), or a crystal in asymmetric Bragg
  diffraction (b) as dispersing element DE. Other components include
  radiation source S, collimating and focusing mirrors M$_{\indrm{C}}$
  and M$_{\indrm{F}}$, and position sensitive detector Det.}
\label{fig001}
\end{figure}

\CT\ \cite{CT30} grating spectrographs are nowadays standard in
infrared, visible, and ultraviolet spectroscopies \cite{SMD64,
  LTR10}. In its classical arrangement, the \CT\ spectrographs
comprise a few elements shown schematically in Fig.~1:
first, a collimating mirror M$_{\indrm{C}}$, which collects photons
from a radiation source S and makes the photon beam parallel,
secondly, a dispersing element DE such as diffraction grating or a
prism, which disperses photons of different photon energies into
different directions, thirdly, a curved mirror M$_{\indrm{F}}$ which
focuses photons of different energies into different locations $F(E)$,
and, last but not least, a spatially sensitive photon detector Det
placed in the focal plane to record the whole photon spectrum.  To
achieve high resolution, what matters most, is the magnitude of the
angular dispersion rate $\dirate=\delta\theta^{\prime}/\delta E$,
which measures the variation of the reflection angle $\theta^{\prime}$
with photon energy $E$  upon reflection from the dispersing element.  For a
given path-length $\pathlength$ (${\mathrm {DE}} \rightarrow {\mathrm
  M}_{\indrm{F}}\rightarrow {\mathrm {Det}}$) from the dispersing
element to the source image $F(E)$ on the detector, the angular
dispersion rate $\dirate$ determines the magnitude of the source
image position variation with photon energy $\delta F(E)=\dirate\pathlength\, \delta E$. In the following, $\dirate\pathlength$
is termed spatial dispersion. The smallest spectral interval $\Delta
E$ between spectral components, which can be resolved is therefore
\begin{equation}
\Delta E\,=\, \frac{1}{\dirate}\,  \frac{\Delta F}{\pathlength },
\label{eq003}
\end{equation}
where $\Delta F$ is the smallest of the two values, either source S
image size on the detector for a particular monochromatic component,
or detector spatial resolution.

Nowadays, diffraction grating manufacturing technology has advanced to
the extent, that grating spectrographs are being successfully used
with much shorter wavelengths, in particular in soft x-ray regime
($\lesssim 1$~keV) \cite{GPD06,BBB10} attaining resolving power of 
$E/\Delta E \simeq 10^4$. Extension into the hard x-ray
regime is, however, not trivial, because of the lack of hard x-ray
optics elements with sufficiently large dispersion rate.

\begin{figure*}
\setlength{\unitlength}{\textwidth}
\begin{picture}(1,0.44)(0,0)
\put(0.0,0.00){\includegraphics[width=1.0\textwidth]{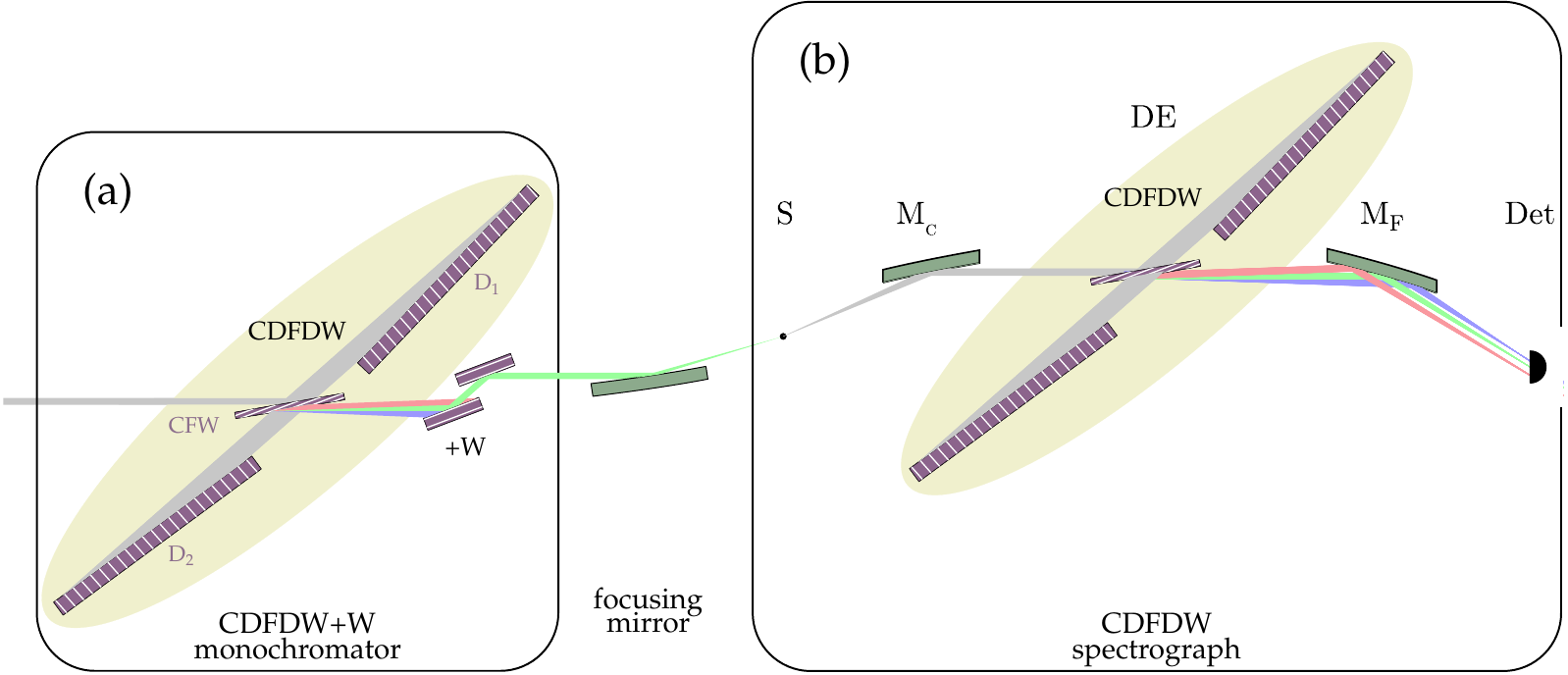}}
\end{picture}
\caption{Layout of an inelastic x-ray scattering (IXS) instrument
  consisting of an x-ray monochromator (a), focusing mirror, and an
  x-ray spectrograph (b). The x-ray monochromator (a) consists of a
  \cddw\ three-crystal (\cw , D$_1$, D$_2$) optics and supplementary
  wavelength selector +W.  The x-ray spectrograph (b), of \CT\ type,
  uses the \cddw\ optics as dispersing element DE (compare
  Fig.~1).}
\label{fig002}
\end{figure*}

A hard x-ray equivalent of the optical diffraction grating is an
asymmetrically cut crystal with reflecting atomic planes at non-zero
asymmetry angle $\eta\not = 0$ to the crystal face, as shown
schematically in Fig.~1(b). In Bragg diffraction from
asymmetrically cut crystals angular dispersion of x-rays takes place
\cite{KB76,BSS95,Shvydko-SB,SLK06}. A collimated x-ray beam at a
glancing angle of incidence $\theta$ to the reflecting atomic planes,
is fanned-out upon reflection with photons of different energies
propagating at different angles $\theta^{\prime}(E)\not = \theta$,
with a dispersion rate
\begin{equation}
  \dirate\, =\, \frac{2}{E}\,\frac{\sin\theta \sin\eta}{\sin(\theta-\eta)} \,=\, \frac{|b|+1}{E}  \tan\theta  \,
  \xrightarrow{\,\,\theta \rightarrow 90^{\circ}\,\,}\, 
  \frac{2\tan\eta}{E}.
\label{eq001}
\end{equation}
Here,  
$b=-\sin(\theta+\eta)/\sin(\theta-\eta)$ is the asymmetry parameter.
Dispersion rates of $\simeq 8-12~\mu$rad/meV were demonstrated in
Bragg diffraction from strongly asymmetrically cut crystals
($\theta-\eta \simeq 2^{\circ}$) close to exact back scattering
($\theta \simeq 90^{\circ}$) where $\dirate $ is maximal
\cite{SLK06,ShSS11}.

The effect of angular dispersion can be used to monochromatize x-rays
beyond the limits determined by the spectral width of Bragg
reflections \cite{Shvydko-SB, SLK06}. This can be accomplished by
using a CDW type optics with three key elements: a collimator (C),
dispersing element (D) and wavelength selector (W) \cite{Shvydko-SB,
  SKR06}. Asymmetrically cut crystals are used as C, D, and W
elements. In particular, demonstrated recently an advanced CDW scheme,
\cddw\ monochromators \cite{ShSS11}, can monochromatize to bandwidths
$\Delta E_{\indrm{\mono}}=0.5$~meV and less, achieving a resolving
power of $E/\Delta E_{\indrm{\mono}}\gtrsim 10^{7}$ at $E=9.1$~keV.

The use of asymmetrically cut crystal as dispersing element was
proposed for a ``focusing monochromator"\cite{KCR09}. The crystal is combined
with a focusing x-ray lens which Fourier transforms angular dispersion
into spatial dispersion. As was correctly noticed in \cite{KCR09} such
optics can be used as a monochromator, however, not as a spectral
analyzer (or spectrograph), as a small angular size of the radiation
source is required for its realization.

As Fig.~1 suggests, the x-ray spectrograph can be realized
by substituting the diffraction grating in the \CT\ scheme by an
asymmetrically cut crystal. However, the mentioned above practically
achievable dispersion rate in single Bragg reflection $\dirate \simeq
10~\mu$rad/meV is barely useful, since only with path lengths
$\pathlength \gtrsim 10$~m a reasonably large spatial dispersion rate
of $\pathlength \dirate \gtrsim 100~\mu$m/meV can be attained.  Here
we show, on the example of CDW x-ray optics
\cite{Shvydko-SB,SLK06,ShSS11}, that multi-crystal arrangements may
feature cumulative angular dispersion rates more than an order of
magnitude larger than those attainable in single Bragg reflections -
Eq.~\ref{eq001}.  It is this enhanced cumulative angular dispersion
rate that makes hard x-ray spectrographs feasible.

\begin{figure*}
\setlength{\unitlength}{\textwidth}
\begin{picture}(1,0.24)(0,0)
\put(0.05,0.00){\includegraphics[width=0.9\textwidth]{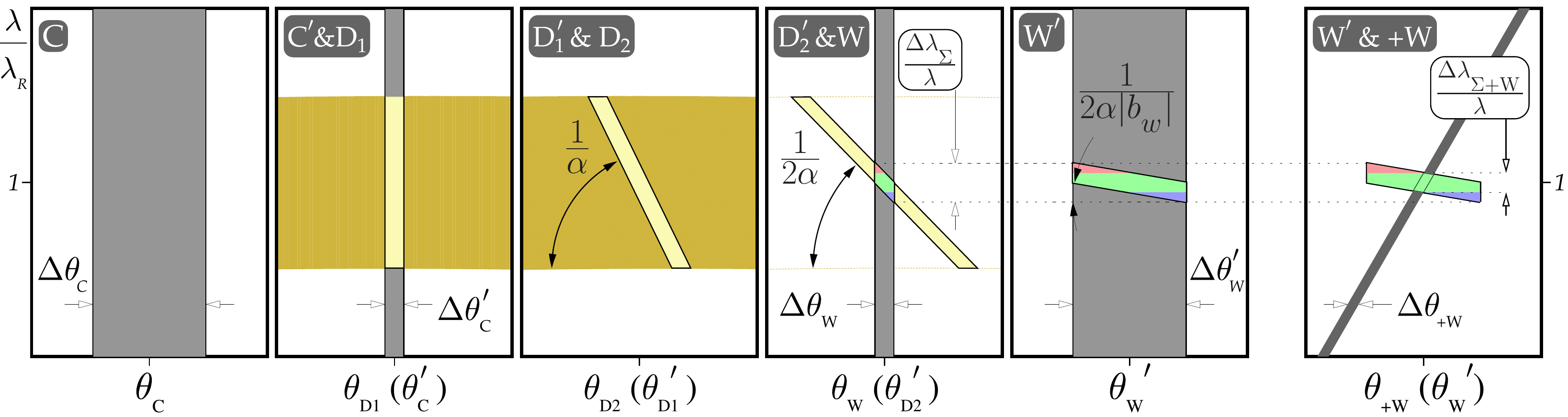}}
\end{picture}
\caption{DuMond diagrams for the sequence of Bragg reflections from
  elements of the \cddw +W optics - Fig.~2(a).  Gray and
  brown stripes are regions of Bragg reflections in the space of x-ray
  wavelengths $\lambda$ and angles of incidence $\theta_{\indrm{H}}$
  or reflection $\theta_{\indrm{H}}^{\prime}$ from the optical
  elements (H=C, D$_1$, D$_2$, W, and +W).  D-element is set into
  backscattering ($\theta_{\indrm{D}}\rightarrow\pi/2$) with the
  center of the reflection spectral range at $\lambda_{\indrm{R}}$.
  Yellow stripes display the overlapping reflection regions of the C-
  and D-elements. Tricolor tetragons display the reflection region
  common for all elements of the \cddw\ optics. Green tetragon in W$^{\prime}$\&+W shows reflection region
  common for all elements of the \cddw +W optics }
\label{fig003}
\end{figure*}

We will use the \cddw\ optics \cite{ShSS11}, an advanced CDW scheme,
as an example, to demonstrate how the enhanced cumulative angular
dispersion rate appears in multi-crystal arrangements.  The \cddw\
optics, shown schematically inside the shaded oval area of
Fig.~2(a), consists of three crystals: \cw , D$_1$, and
D$_2$ executing five successive Bragg reflections.  Each of the five
reflection has its own key function termed as C (collimator), D
(dispersing element), \filter\ (anomalous transmission filter), D, and
W (wavelength selector) respectively. For a detailed discussion of the
principles of the \cddw\ optics, we refer to \cite{ShSS11}.  Here, we
use DuMond diagram analysis \cite{DuMond37} to calculate the angular
acceptance $\Delta \theta_{\indrm{\mono }}$ for the input radiation,
angular spread $\Delta \theta_{\indrm{\mono }}^{\prime}$ and spectral
bandwidth for the output radiation $\Delta E_{\indrm{\mono }}$, but
most importantly the dispersion rate after each Bragg reflection, and
the cumulative dispersion rate $\diratecum $ of entire optics.
Figure~3 shows DuMond diagrams for the sequence of all
Bragg reflections, except for the anomalous transmission through the
\cw\ crystal (\filter -reflection). This omission does not change
substantially main results of the analysis.

Gray and brown stripes are the regions of Bragg reflections in the
space of x-ray wavelengths $\lambda$ and angles of incidence
$\theta_{\indrm{H}}$ or reflection $\theta_{\indrm{H}}^{\prime}$ from
the crystals in one of the reflections H=C, D$_1$, D$_2$, or W. All
the crystals are asymmetrically cut with asymmetry angle
$\eta_{\indrm{H}}$.  The yellow stripes display the overlapping
reflection regions of the C- and D-crystals. Tricolor tetragons
display the reflection region common for all elements.  D-element is
set into backscattering ($\theta_{\indrm{D}}\rightarrow\pi/2$) with
the center of the reflection spectral range at $\lambda_{\indrm{R}}$.

The asymmetry parameter of the \cw\ crystal in the C-reflection, has
to be chosen small $|b_{\indrm{C}}|\ll 1$ to accept incident photons
in a broad angular range $\Delta \theta_{\indrm{C }}$ and collimate
them into a beam with a small divergence $\Delta
\theta_{\indrm{C}}^{\prime}$:
\begin{equation} 
  \Delta \theta_{\indrm{C }}  \,=\, \dai{\mathrm C}/ \sqrt{|b_{\indrm{C}}|},\hspace{0.5cm}
  \Delta
  \theta_{\indrm{C}}^{\prime} \,=\, \dai{\mathrm C}/
  \sqrt{|b_{\indrm{C}}|}.
\label{eq002a}
\end{equation}
For a low indexed Bragg reflection, such as Si(220), the angular width
in symmetric reflection geometry $\dai{\mathrm C} \simeq 23~\mu$rad.
By choosing $|b_{\indrm{C}}|= 1/21.5$, as in \cite{ShSS11}, the
angular acceptance of the C-element, as well as of the entire \cddw\
optics, becomes large $\Delta \theta_{\indrm{C }} = 106~\mu$rad, while
the angular spread of x-ray impinging on D$_1$ crystal - small:
$\Delta \theta_{\indrm{C}}^{\prime}= 5~\mu$rad. The phase space for
C-reflection is shown in diagrams C and C$^{\prime}$\&D$_1$, as stripes
almost vertical due to a small Bragg angle $\simeq
20.8^{\circ}$. Yellow strip indicates the common phase space for
photons upon both reflections C- and D$_1$. In diagram
D$_1^{\prime}$\&D$_2$, it is shown rotated by an angle
$\arctan(1/\alpha)$, where $\alpha=2\tan\eta_{\indrm{D}}$ -
Eq.~\eqref{eq001} - due to angular dispersion upon reflection
D$_1^{\prime}$.  Reflection D$_2$ doubles the dispersion rate to
$4\tan\eta_{\indrm{D}}$, indicated by the larger inclination of the
phase space in diagram D$_2^{\prime}$\&W. Only those photons are
reflected from the \cw\ crystal in the W-reflection, which are
concentrated in the small angular range of the W-reflection $\Delta
\theta_{\indrm{W}} \,=\, \dai{\mathrm W}/ \sqrt{|b_{\indrm{W}}|} =
\Delta \theta_{\indrm{C}}^{\prime} $ . Here
$|b_{\indrm{W}}|=1/|b_{\indrm{C}}|=21.5 \gg 1$. The W-reflection
selects photons in a small angular range, and most importantly in a
bandwidth much smaller than the bandwidth of the D-reflections:
\begin{equation}
\frac{\Delta E_{\indrm{\mono }}}{E} \,=\,\frac{\Delta \lambda_{\indrm{\mono }}}{\lambda}\, =\, \frac{\Delta\theta_{\indrm{C}}^{\prime} +\Delta\theta_{\indrm{W}}}{4\tan\eta_{\indrm{D}}}=\, \frac{\Delta\theta_{\indrm{W}}}{2\tan\eta_{\indrm{D}}}.
\label{eq002b}
\end{equation}
Remarkably, the photons are spread upon the W-reflection into a large
angle, which is 
equal to the angular acceptance of the \cddw\ optics
\begin{equation}
\Delta \theta_{\indrm{\mono }}^{\prime}\, =\Delta \theta_{\indrm{\mono }}\, =\, \dai{\mathrm W} \sqrt{|b_{\indrm{W}}|}=\, \dai{\mathrm C}/ \sqrt{|b_{\indrm{C}}|},  
\label{eq002c}
\end{equation}
within the accuracy of the DuMond diagram analysis.

\begin{figure}
\setlength{\unitlength}{\textwidth}
\begin{picture}(1,0.51)(0,0)
\put(0.0,0.0){\includegraphics[width=0.5\textwidth]{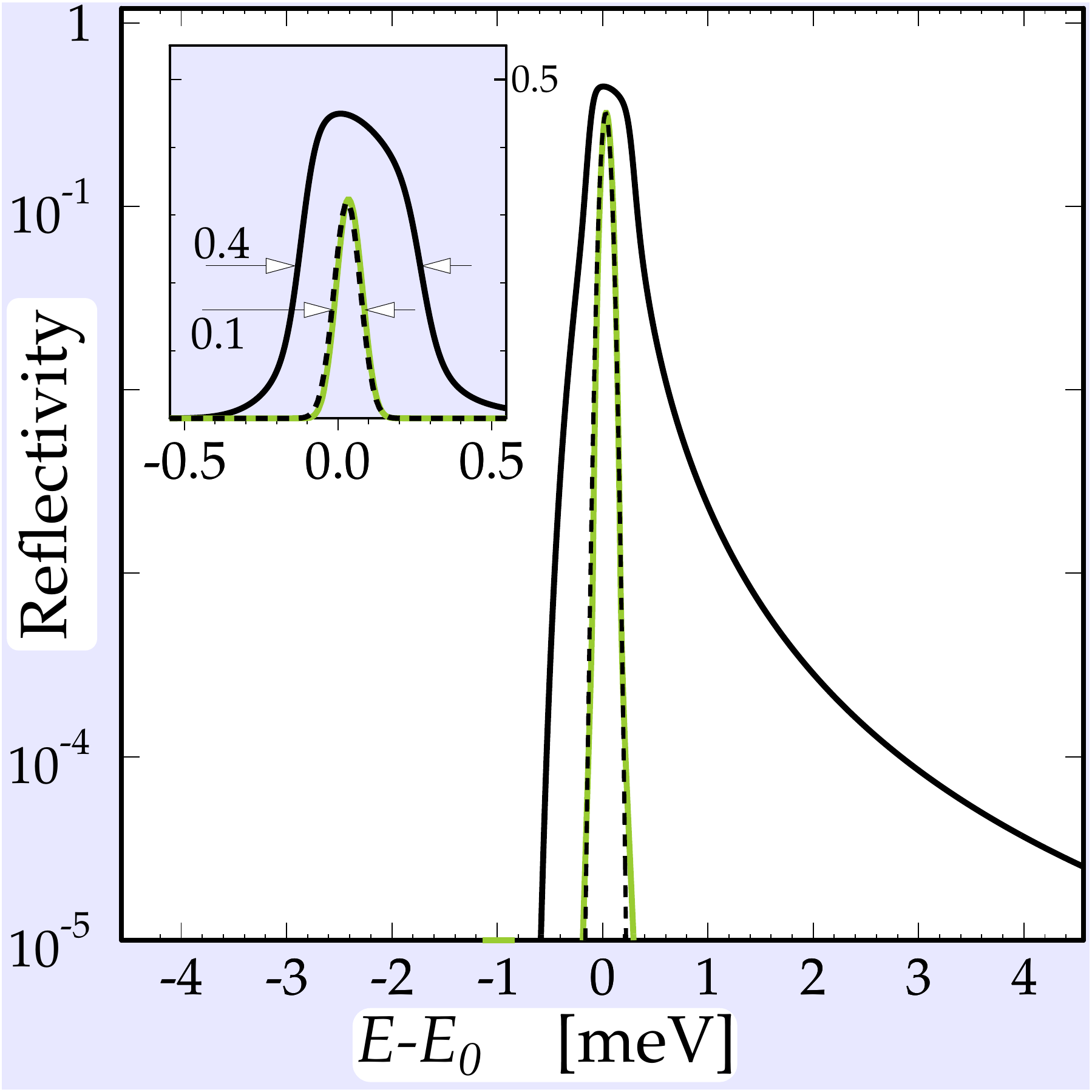}}
\end{picture}
\caption{Dynamical theory calculations of the spectral distribution of
  x~rays after five successive reflection in the \cddw\ optics (black
  line) and after additional two reflections from the supplementary
  wavelength selector +W (green line). Black dashed line shows
  Gaussian distribution of the same full width at half maximum. Inset
  shows the same distributions on the linear scale. The Bragg
  reflections and crystal parameters are used for the \cddw\ optics
  the same as in \cite{ShSS11}. The Si(220) reflection in symmetric
  geometry is chosen for +W.}
\label{fig004}
\end{figure}

Here we are arriving at the crucial point. From diagram W$^{\prime}$
in Fig.~3, it follows, that the W-reflection not only
increases the beam divergence, it also increases by the same factor of
$|b_{\indrm{W}}|$ the tangent of the phase space, i.e. it increases
the angular dispersion rate from $2\dirate$ to
\begin{equation}\diratecum\, =\,2 \dirate\,
  |b_{\indrm{W}}|= \frac{4\tan\eta_{\indrm{D}}}{E}\, |b_{\indrm{W}}| .
\label{eq005}
\end{equation}
That means that the cumulative angular dispersion rate $\diratecum$,
the \cddw\ optics endowed with, is by a factor of $\simeq
|b_{\indrm{W}}|$ greater than the largest dispersion rate $\dirate$
achievable in single Bragg reflection. In the example of
\cite{ShSS11}, $\eta_{\indrm{D}}=88^{\circ}$, $\dirate =
6.3~\mu$rad/meV, $|b_{\indrm{W}}|\simeq 21.5$, and therefore
$\diratecum = 270~\mu$rad/meV. The dispersion rate enhancement is
large $\diratecum/\dirate \gtrsim 40$,
and can be used at least twofold.

First, the bandwidth of the \cddw\ monochromator can be further
reduced by applying one more Bragg reflection with an angular width
$\Delta \theta_{\indrm{+W}} \ll \Delta \theta_{\indrm{W}}^{\prime}$,
see Fig.~3(W$^{\prime}$\&+W). To keep the beam direction
unchanged, a better choice is two equivalent Bragg reflections from
crystals in non-dispersive configuration, as shown schematically in
Fig.~2(a), and denoted by +W for supplementary wavelength
selector.  Figure~4 shows dynamical theory calculations for
+W as Si crystals in the (220) Bragg reflection in symmetric geometry,
other parameters of the \cddw\ optics equivalent to those realized in
\cite{ShSS11}, and a $20~\mu$rad angular spread for incident photons.
The application of the supplementary wavelength selector reduces the
bandwidth from 0.4~meV to 0.1~meV, what correlates with the estimation
using the DuMond diagram W$^{\prime}$\&+W.  However, not only the
+W-reflections reduce the bandwidth, they make the wings of the
spectral distribution extremely sharp, as sharp as of the Gaussian
distribution. With smaller $\Delta \theta_{\indrm{+W}}$ the bandwidth
can be further reduced.

Secondly, the large dispersion rate enhancement can be used to realize
a hard x-ray spectrograph.  Its scheme is shown in
Fig.~2(b), as a part of an IXS instrument, comprising also
the \cddw +W monochromator and a 2D-mirror focusing x-rays on sample
S.  The \cddw\ spectrograph is similar to the \CT\ spectrograph
presented in Fig.~1, with one major difference.  The
diffraction grating (or the single crystal) is substituted by a
multi-crystal optics with enhanced dispersion rate. The components of
the spectrograph are also almost equivalent to the components of the
CDW analyzer proposed in \cite{Shvydko-SB} (Sec.~6.3).
The crucial difference is that the 2D-focusing mirror M$_{\indrm{F}}$
is added downstream the multi-crystal optics. This converts the
analyzer into a spectrograph with potentially much higher spectral
resolution. If mirrors M$_{\indrm{C}}$ and M$_{\indrm{F}}$ are
equivalent and perfect, the system produces 1:1 image of the source S
(x-ray scattered from sample S) on the detector Det with monochromatic
x-rays independent of the distances between the mirrors and the
distances of the \cddw\ optics to the mirrors. On the other hand, the
path-length $\pathlength$ from the \cddw\ optics to the source image
scales the linear dispersion rate $\diratecum\pathlength$.

The cumulative dispersion rate of the \cddw\ optics studied in
\cite{ShSS11} is $\diratecum= 270~\mu$rad/meV. This results in a large
linear dispersion rate $\dirate_{\indrm{\mono }} \pathlength=
270~\mu$m/meV, even with a reasonably small $\pathlength\simeq
1$~m. $\pathlength$ can be increased without degrading the source
image size $\Delta F$. The resolution of the spectrograph is limited,
according to Eq.~\eqref{eq003} by $\Delta F$.  Assuming the sources
size to be $\simeq 5-10~\mu$rad, as discussed in \cite{Shvydko-SB}
(Sec.~6.3), $\Delta F\lesssim 50~\mu$m is feasible, also if some
broadening due to figure errors in mirrors are involved.  Therefore
$\Delta E_{\indrm{\mono}}\lesssim 0.1$~meV, and resolving power
$E/\Delta E_{\indrm{\mono}} \gtrsim 10^{8}$ is feasible.

The DuMond diagram analysis presented in Fig.~3 suggests,
that the enhanced cumulative dispersion rate appears in multi-crystal
optics if a wavelength selector is applied as asymmetrically cut
crystal with $|b_{\indrm{W}}|\gg 1$. This property is general. It is
not a unique feature of the CDW-type crystal optics.
The \mmixed\ monochromator composed of four asymmetrically cut
crystals \cite{Yabashi01}, is another example. Using the results of
the DuMond diagram analysis presented in Sec.~3.5 of \cite{Shvydko-SB}
by Eqs.~(3.32)-(3.33), it is easy to show that the cumulative angular
dispersion rate of the \mmixed\ monochromator $\diratecum \propto
\Delta\theta^{\prime}/\Delta E \simeq \tan\theta
|b_{\ind{3}}b_{\ind{4}}|/E$, where $|b_{\ind{n}}| \gg 1$ is the
absolute value of the asymmetry parameter for the 3rd and 4th Bragg
reflection. The dispersion rate for a single reflection is
$\dirate\propto\tan\theta\, (|b_{\ind{3}}|+1)/E$, according to
Eq.~\eqref{eq001}. Therefore, cumulative $\diratecum$ is enhanced by a
factor of $|b_{\ind{4}}|$.

In conclusion, multi-crystal x-ray optics with crystals in asymmetric
diffraction may exhibit enhanced cumulative dispersion rate, more than
an order of magnitude larger than the dispersion rate associated with
a single Bragg reflection. Such optics can be used as dispersing
element in \CT -type x-ray spectrographs.  \CT -type spectrograph with
CDW dispersing optics may feature resolving power beyond $10^{8}$. \CT
scheme could be also used to realized high-resolution spectrographs in
soft x-ray regime.

S.~Stoupin is acknowledged for reading the manuscript and valuable
suggestions. Work was supported by the U.S. Department of Energy,
Office of Science, Office of Basic Energy Sciences, under Contract
No. DE-AC02-06CH11357.


\begin{thebibliography}{10}
\expandafter\ifx\csname url\endcsname\relax
  \def\url#1{\texttt{#1}}\fi
\expandafter\ifx\csname urlprefix\endcsname\relax\def\urlprefix{URL }\fi
\providecommand{\bibinfo}[2]{#2}
\providecommand{\eprint}[2][]{\url{#2}}

\bibitem{Shvydko-SB}
\bibinfo{author}{Shvyd'ko, Yu.}
\newblock \emph{\bibinfo{title}{X-Ray Optics -- High-Energy-Resolution
  Applications}}, vol.~\bibinfo{volume}{98} of \emph{\bibinfo{series}{Optical
  Sciences}} (\bibinfo{publisher}{Springer}, \bibinfo{address}{Berlin
  Heidelberg New~York}, \bibinfo{year}{2004}).

\bibitem{SLK06}
\bibinfo{author}{Shvyd'ko, Yu.~V.} \emph{et~al.}
\newblock \bibinfo{title}{Bragg diffraction of x~rays in asymmetric
  backscattering geometry}.
\newblock \emph{\bibinfo{journal}{Phys. Rev. Lett.}}
  \textbf{\bibinfo{volume}{97}}, \bibinfo{pages}{235502}
  (\bibinfo{year}{2006}).

\bibitem{ShSS11}
\bibinfo{author}{Shvyd'ko, Yu.}, \bibinfo{author}{Stoupin, S.},
  \bibinfo{author}{Shu, D.} \& \bibinfo{author}{Khachatryan, R.}
\newblock \bibinfo{title}{Using angular dispersion and anomalous transmission to shape
  monochromatic x rays}.
\newblock \emph{\bibinfo{journal}{Phys. Rev. A (accepted), eprint arXiv:1108.2487}}
   (\bibinfo{year}{2011}).

\bibitem{CT30}
\bibinfo{author}{Czerny, M.} \& \bibinfo{author}{Turner, A.~F.}
\newblock \bibinfo{title}{{\"{U}}ber den {Astigmatismus} bei
  {Spiegelspektrometern}}.
\newblock \emph{\bibinfo{journal}{Z. f. Physik}} \textbf{\bibinfo{volume}{61}},
  \bibinfo{pages}{792--797} (\bibinfo{year}{1930}).

\bibitem{SMD64}
\bibinfo{author}{Shafer, A.~B.}, \bibinfo{author}{Megill, L.~R.} \&
  \bibinfo{author}{Droppelman, L.}
\newblock \bibinfo{title}{Optimization of the {Czerny-Turner} {S}pectrometer}.
\newblock \emph{\bibinfo{journal}{J. Opt. Soc. Am.}}
  \textbf{\bibinfo{volume}{54}}, \bibinfo{pages}{879--886}
  (\bibinfo{year}{1964}).

\bibitem{LTR10}
\bibinfo{author}{Lee, K.-S.}, \bibinfo{author}{Thompson, K.~P.} \&
  \bibinfo{author}{Rolland, J.~P.}
\newblock \bibinfo{title}{Broadband astigmatism-corrected {Czerny-Turner}
  spectrometer}.
\newblock \emph{\bibinfo{journal}{Opt. Express}} \textbf{\bibinfo{volume}{18}},
  \bibinfo{pages}{23378 -- 23384} (\bibinfo{year}{2010}).

\bibitem{GPD06}
\bibinfo{author}{Ghiringhelli, G.} \emph{et~al.}
\newblock \bibinfo{title}{{SAXES}, a high resolution spectrometer for resonant
  x-ray emission in the 400-1600 {eV} energy range}.
\newblock \emph{\bibinfo{journal}{Rev. Sci. Instrum.}}
  \textbf{\bibinfo{volume}{77}}, \bibinfo{pages}{113108}
  (\bibinfo{year}{2006}).

\bibitem{BBB10}
\bibinfo{author}{Braicovich, L.} \emph{et~al.}
\newblock \bibinfo{title}{Magnetic excitations and phase separation in the
  underdoped {La}$_{2-x}${Sr}$_{x}${CuO}$_{4}$ superconductor measured by
  resonant inelastic x-ray scattering}.
\newblock \emph{\bibinfo{journal}{Phys. Rev. Lett.}}
  \textbf{\bibinfo{volume}{104}}, \bibinfo{pages}{077002}
  (\bibinfo{year}{2010}).

\bibitem{KB76}
\bibinfo{author}{Kuriyama, M.} \& \bibinfo{author}{Boettinger, W.~J.}
\newblock \bibinfo{title}{On the angular divergence of out-going beams in an
  asymmetric diffraction geometry}.
\newblock \emph{\bibinfo{journal}{Acta Cryst.}} \textbf{\bibinfo{volume}{A32}},
  \bibinfo{pages}{511} (\bibinfo{year}{1976}).

\bibitem{BSS95}
\bibinfo{author}{Brauer, S.}, \bibinfo{author}{Stephenson, G.} \&
  \bibinfo{author}{Sutton, M.}
\newblock \bibinfo{title}{Perfect crystals in the asymmetric bragg geometry as
  optical elements for coherent x-ray beams}.
\newblock \emph{\bibinfo{journal}{J. Synchrotron Rad.}}
  \textbf{\bibinfo{volume}{2}}, \bibinfo{pages}{163--173}
  (\bibinfo{year}{1995}).

\bibitem{SKR06}
\bibinfo{author}{Shvyd'ko, Yu.~V.} \emph{et~al.}
\newblock \bibinfo{title}{Progress in the development of new optics for very
  high resolution inelastic x-ray scattering spectroscopy}.
\newblock \emph{\bibinfo{journal}{AIP Conf. Proc.}}
  \textbf{\bibinfo{volume}{879}}, \bibinfo{pages}{737--745}
  (\bibinfo{year}{2007}).

\bibitem{KCR09}
\bibinfo{author}{Kohn, V.~G.}, \bibinfo{author}{Chumakov, A.~I.} \&
  \bibinfo{author}{R{\"{u}}ffer, R.}
\newblock \bibinfo{title}{{Wave theory of focusing monochromator of synchrotron
  radiation}}.
\newblock \emph{\bibinfo{journal}{J. Synchrotron Rad.}}
  \textbf{\bibinfo{volume}{16}}, \bibinfo{pages}{635--641}
  (\bibinfo{year}{2009}).

\bibitem{DuMond37}
\bibinfo{author}{{DuMond}, J. W.~M.}
\newblock \bibinfo{title}{Theory of the use of more than two successive x-ray
  crystal reflections to obtain increased resolving power}.
\newblock \emph{\bibinfo{journal}{Phys. Rev.}} \textbf{\bibinfo{volume}{52}},
  \bibinfo{pages}{872--883} (\bibinfo{year}{1937}).

\bibitem{Yabashi01}
\bibinfo{author}{Yabashi, M.}, \bibinfo{author}{Tamasaku, K.},
  \bibinfo{author}{Kikuta, S.} \& \bibinfo{author}{Ishikawa, T.}
\newblock \bibinfo{title}{An x-ray monochromator with an energy resolution of
  $8\times 10^{-9}$ at 14.4 {keV}}.
\newblock \emph{\bibinfo{journal}{Rev. Sci. Instrum.}}
  \textbf{\bibinfo{volume}{72}}, \bibinfo{pages}{4080} (\bibinfo{year}{2001}).

\end{thebibliography}



\end{document}